# Correlations of gas, dust and young stellar populations in the spiral galaxy NGC 7331

**Selçuk TOPAL[1,*]**

[1] *Yuzuncu Yil University, Faculty of Science, Department of Physics, Van, Turkey.*

## Abstract

*Molecular clouds (MCs) in galaxies are complex places with many phases. It is, therefore, essential to study the physics and kinematics of the MCs using multiple emission lines. We probe the physics of the molecular gas and dust in the nearby spiral galaxy NGC 7331 using multiple emission lines, i.e. carbon monoxide (CO), 24μm and far-ultraviolet (FUV) data. 14 positions were targeted across the gaseous disc of NGC 7331. We found that CO intensities, gas mass, gas surface density, and 24μm-to-FUV flux ratio (i.e. the extinction) increase up to about 40 arcsec from the centre and then start to decrease. There is a positive correlation between most of the pair of parameters studied (except FUV flux density). The beam-averaged physical parameters on the eastern side of the disc show higher median values than those on the western side. Our results indicate that the star formation activity, stellar populations and overall physical properties of the ISM are different on either side of the disc. Our study provides notable insights into the complex nature of the interstellar medium (ISM) in galaxies and has the potential to provoke future higher-resolution studies yet to come.*

*Keywords: Galaxies, spiral galaxies, molecular gas, far-ultraviolet emission, 24μm emission*

## NGC 7331 sarmal gökadasında gaz, toz ve genç yıldız popülasyonları arasındaki korelasyon

## Öz

*Galaksilerdeki moleküler bulutlar (MC'ler), birçok evresi olan karmaşık yapılardır. Bu nedenle, çoklu salma çizgileri kullanarak MC'lerin fiziğini ve kinematiğini incelemek önemlidir. Bu çalışmada, karbon monoksit (CO), 24 μm ve uzak moröte (FUV) salma çizgilerini kullanarak bize görece yakın NGC 7331 sarmal gökadasındaki moleküler*

---
*Selçuk TOPAL, selcuktopal@yyu.edu.tr, https://orcid.org/0000-0003-2132-5632

*gaz ve tozun fiziğini araştırıyoruz. NGC 7331 gökadasının gazlı diski boyunca 14 konum hedeflendi. CO yoğunluklarının, gaz kütlesinin, gaz yüzey yoğunluğunun ve 24 µm / FUV akı oranının (yani sönümleme) merkezden yaklaşık 40 yay saniyeye kadar arttığını ve sonra azalmaya başladığını bulduk. Çalışılan parametre çiftlerinin çoğu arasında pozitif bir korelasyon bulunmaktadır (FUV akı yoğunluğu hariç). Diskin doğu tarafında elde edilen fiziksel parametreler batı tarafındakilerden daha yüksek medyan değerleri göstermektedir. Sonuçlar, yıldız oluşum aktivitesinin, yıldız popülasyonlarının ve ISM'in genel fiziksel özelliklerinin diskin her iki tarafında farklı olduğunu göstermektedir. Çalışmamız, galaksilerdeki ISM'in karmaşık doğası hakkında dikkate değer bilgiler sağlamakta ve gelecekte yapılması planlanan daha yüksek çözünürlüklü çalışmaları teşvik etme potansiyeline sahiptir.*

***Anahtar kelimeler:*** *Gökadalar, sarmal gökadalar, moleküler gaz, uzak morötesi salma, 24 µm salma*

## 1. Introduction

Studying multiple emission features in the interstellar medium (ISM) of galaxies, such as molecular, far-infrared (FIR) and far-ultraviolet (FUV) emissions, provides us with an opportunity to study the formation and evolution of stars in greater details. Carbon monoxide (CO) is a particularly useful tracer for the gas clouds since $^{12}$CO(1-0) transition lies only about 5.5 K from the ground state and it is easier to detect compared to hydrogen molecule ($H_2$). CO is therefore easily detectable in cold molecular clouds with an average temperature of 10 K. $^{12}$CO(1-0) is also widely used to estimate the total molecular gas budget in galaxies [1]. Spiral galaxies host a considerable amount of molecular gas, dust and young stellar populations compared to elliptical galaxies which are poor in gas and dust. The spirals are therefore ideal targets to study star formation processes in galaxies.

In molecular clouds (MCs) various phenomena are at play affecting the physics and chemistry of the clouds simultaneously. Young and massive stars could affect the physics and chemistry of the ISM through FUV radiation and stellar winds [2]. FUV radiation and dust play a vital role in the formation and destruction of $H_2$ [3,4] and also in regulating the CO abundances in the ISM [5]. It is well known that star formation is usually correlated with dust emission heated by young massive stars through their strong FUV radiation [6]. Dust emission at 24µm is particularly important since it traces star formation rate in galaxies [7]. Given the strong connection between 24µm and FUV radiation, the ratio of 24µm-to-FUV flux densities could give insights on the extinction and star formation efficiency and also shows correlation with some other physical properties, such as molecular gas mass, metallicity, luminosity and distance from the galactic centre [8].

It is necessary to use multi-wavelength data to probe not only the correlations of gas, dust and stars but also the physics and kinematics driving the ISM across the whole disc of a galaxy better. We targeted NGC 7331 for several reasons. First of all, it has multitude of data sets from FUV to mm wavelengths. Secondly, the galaxy has a suitable position angle allowing us to study the ring-like gas distribution. Thirdly, it is bright in CO, and last but not least, it is a nearby galaxy providing us with a better linear resolution across the disc of the galaxy. NGC 7331 is, therefore, an impeccable target to serve the main purposes of this study. We target multiple positions across the ring-like gaseous disc of NGC 7331 to probe the complex nature of the ISM at each selected position by following a multi-wavelength approach, for the first time for this galaxy. We aim to answer some outstanding questions about the life cycle of the gas clouds, stars and the evolution of galaxies. Those questions are as follows. How is molecular gas (i.e. the fuel for star formation) correlated with dust emission, extinction and young stellar populations? How do young stars affect the overall physics of the ISM? What are the differences in the physical properties at different locations over the disc of a spiral galaxy? For the first time for this galaxy, we aim to understand the interplay of gas, dust and young stellar populations using the multiple emission lines obtained at multiple positions over the disc of NGC 7331. NGC 7331 is located at a distance of 14.5 Mpc (*HyperLEDA* database; [9]). The basic properties of this galaxy are given in Table 1.

We describe the literature data and analysis in Section 2 and present the results and related discussions in Section 3. Finally, Section 4 briefly summarises our main conclusions.

Table 1. The basic properties of NGC 7331.

| Property | Value | Ref. |
|---|---|---|
| Galaxy type | SA | a |
| RA (J2000) | $22^h37^m04.014^s$ | a |
| DEC (J2000) | $+34^d24^m55.87^s$ | a |
| Major-axis diameter | 10 arcmin | a |
| Minor-axis diameter | 3 arcmin | a |
| Heliocentric velocity ($V_\odot$) | 816 km/s | a |
| Distance | 14.5 Mpc | b |
| Inclination | 70 degree | b |

(a) NASA Extragalactic Database (NED); (b) *HyperLEDA (*http://leda.univ-lyon1.fr*)*

## 2. Material ve method

### 2.1. *CO, dust and far-ultraviolet data*

Literature CO data for NGC 7331 were obtained from *Berkeley-Illinois-Maryland Association Survey Of Nearby Galaxies* (*BIMA SONG*; [10]). *BIMA SONG* CO map covers an area of 5.7 x 5.6 arcmin$^2$ over the galaxy with a spatial and spectral resolution of 6.1 x 5.4 arcsec$^2$ and 10 kms$^{-1}$, respectively. The data for 24μm and FUV emissions ($\lambda_{eff}$ = 1516 Å) were taken from *Spitzer* and *the Galaxy Evolution Explorer* (*GALEX*; [11]) surveys, respectively. The spatial resolutions for 24μm and FUV data are 6 arcsec and 4 arcsec, respectively. For more details on the observational parameters, we refer the reader to the associated papers.

### 2.2. *Integrated line intensity and moment maps*

Data analysis and imaging were done using *the Multichannel Image Reconstruction Image Analysis and Display* (*MIRIAD*) package [12] and *Interactive Data Language (IDL)* environment. The integrated CO line intensity is defined as $\int T_{mb}dv$, where $T_{mb}$ is the antenna main beam brightness temperature in the unit of Kelvin (K) and $dv$ is the Full-Width at Half-Maximum (FWHM) of the line profile in unit of $km\,s^{-1}$. We calculated the integrated CO line intensities by fitting a single Gaussian to the line profiles using *IDL* and *MPFIT* [13]. *MPFIT* optimises the fit by employing the Levenberg–Marquardt minimization algorithm.

We used the original (i.e. not convolved) CO data cube to create CO moment maps with the highest resolution possible, i.e. integrated intensity map (moment 0), velocity map (moment 1), and velocity dispersion map (moment 2). To create the moment maps, we first specified an adjoining region for the source emission. The method is based on spectral and spatial smoothing of the original data cube. After the smoothing, a $3\sigma$ cutoff (where $\sigma$ is the rms noise in the cube estimated using the channels with no emission) was applied to the (smoothed) cube to obtain the (smoothed) moment maps. The regions for source contiguous emission were defined using *IDL* algorithm *label_region*. The (smoothed) moment maps were then used to create a 3D mask. The 3D mask was applied to the original (unsmoothed) data cube to define the continuous region for the source emission in the cube. We finally obtained the moment maps from the (masked) cube using *MIRIAD* task *moment*. All moment maps are shown in Figure 1.

### 2.3. *Position selection and spectrum extraction*

We defined 14 positions side-by-side along the ring-like gas distribution surrounding the centre of the galaxy. The size of each region was defined by considering the thickness of the gas distribution along the line of sight so that each selected circular region would cover the CO emission properly. Each position was found to be properly covered by a circle of 20 arcsec in size. CO data were convolved to the adopted beam size of 20 arcsec using *MIRIAD* task *convol* before applying further analysis. The projected 14 positions throughout the disc of the galaxy are shown in Figure 2. The $^{12}$CO(1-0) spectrum was extracted from each selected position in the (convolved) data cube using *MIRIAD* task *imspect*. The spectra are shown in Figure A.1 in Appendix.

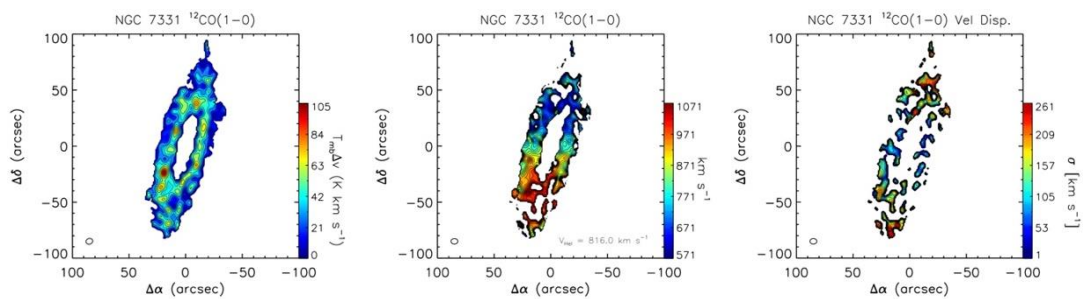

Figure 1. The integrated CO intensity map (moment 0; left), velocity map (moment 1; middle) and velocity dispersion map (moment 2; right) are shown. The moment 0 map has contour levels range from 10% to 100% of the maximum intensity with an increase of 10%. The contour levels on the velocity map and velocity dispersion map are spaced by 20 km/s. The beam size is also shown in the lower-left corner of each panel.

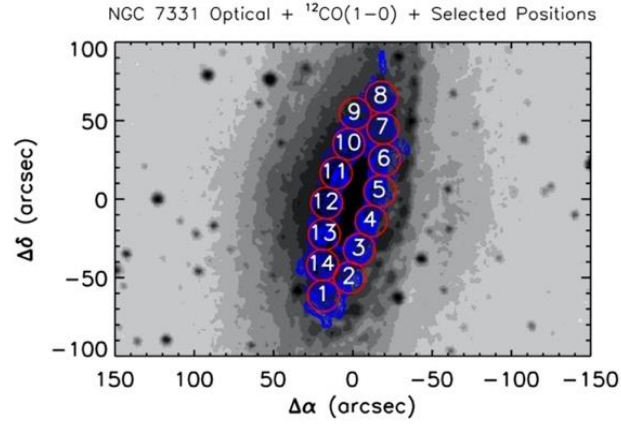

Figure 2. The CO contour map (shown in blue contours) is on the top of an optical image of the galaxy (greyscale contour on the background). Each of 14 projected positions over the CO contour map is shown with a red circle and also indicated with the position number. Each red circle has a diameter of 20 arcsec or equivalently 1.4 kpc at the distance of the galaxy (see Table 1).

### 2.4. *Molecular gas mass and gas surface density*

To calculate the molecular gas mass averaged over the beam, $M_{H_2}$, at each projected position, we adopted a value of $X_{CO} = 2 \times 10^{20}$ cm$^{-2}$ (K km s$^{-1}$)$^{-1}$ for the CO-to-H$_2$ conversion factor, which is suitable for the disc of nearby spiral galaxies [14-16]. The expression used to calculate $M_{H_2}$ is shown below [16].

$$\frac{M_{H_2}}{M_\odot} = 7.2 \times 10^6 \times \left( \frac{I_{1-0}}{\text{K km s}^{-1}} \right) \tag{1}$$

where $I_{1-0}$ is the integrated CO line intensity. The gas surface density, $\sum H_2$, was also estimated (i.e. $\alpha_{CO} = \frac{\sum H_2}{I_{1-0}}$, where $\alpha_{CO} = 3.2\ M_\odot$ (K km s$^{-1}$)$^{-1}$ for the adopted value of $X_{CO}$ [17]). The values of $M_{H_2}$ and $\sum H_2$ are listed in Table 2 and also shown in Figure 3.

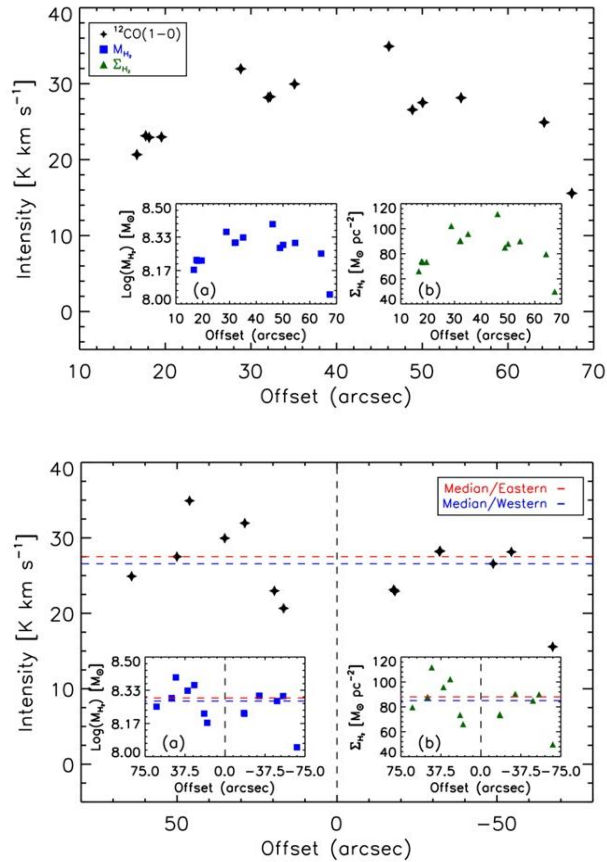

Figure 3. The integrated CO line intensities (black diamonds), molecular gas mass, $M_{H_2}$ (blue squares; embedded panel a), and gas surface density, $\sum H_2$ (green triangles; embedded panel b) are shown as a function of projected distance from the galaxy's centre (top) and on each side of the disc (bottom). The horizontal red dashed and blue dashed lines on the lower panel indicate the median values for the parameters on the eastern and western sides of the disc, respectively. The vertical black dashed lines on the image at the bottom indicate the centre of the galaxy.

Table 2. Total molecular gas mass and gas surface density at each studied position.

| Position | $Log(M_{H_2})$ $[M_\odot \times 10^7]$ | $\sum H_2$ $[M_\odot \text{ pc}^{-2}]$ | Position | $Log(M_{H_2})$ $[M_\odot \times 10^7]$ | $\sum H_2$ $[M_\odot \text{ pc}^{-2}]$ |
|---|---|---|---|---|---|
| 1* | $17.8 \pm 0.2$ | $79.7 \pm 5.5$ | 8 | $11.1 \pm 0.2$ | $49.8 \pm 4.3$ |
| 2* | $19.7 \pm 0.3$ | $88.0 \pm 6.7$ | 9 | $20.1 \pm 0.4$ | $90.0 \pm 9.1$ |
| 3 | $20.2 \pm 0.2$ | $90.5 \pm 4.8$ | 10* | $21.4 \pm 0.3$ | $95.8 \pm 7.4$ |
| 4 | $16.4 \pm 0.2$ | $73.4 \pm 4.3$ | 11* | $16.4 \pm 0.2$ | $73.5 \pm 5.3$ |
| 5 | $16.6 \pm 0.2$ | $74.0 \pm 4.9$ | 12* | $14.8 \pm 0.2$ | $66.1 \pm 4.0$ |
| 6 | $20.2 \pm 0.3$ | $90.2 \pm 7.6$ | 13* | $22.9 \pm 0.3$ | $102.2 \pm 5.6$ |
| 7 | $19.0 \pm 0.3$ | $85.0 \pm 7.3$ | 14* | $24.9 \pm 0.4$ | $111.7 \pm 8.6$ |

*The positions in the eastern side of the disc with respect to the galaxy's centre ($\alpha = 22^h 37^m 04.014^s$ and $\delta = +37^d 24^m 55.87^s$). The rest of the projected positions are located on the western side of the galaxy. See also Figure 1 for the exact locations of the projected positions over the galaxy.

**2.5 *24 µm and FUV data***
We first applied the necessary unit conversion to 24 µm and FUV data to obtain the flux density in the unit of Jansky (Jy). We then calculated the Gaussian weighted beam-averaged total 24 µm and FUV flux densities for each projected position as follows. We first multiplied each pixel in each image by a 2D normalised Gaussian centred at each position with the FWHM equal to 20 arcsec. We then summed all the fluxes in each image to obtain the Gaussian weighted flux density for each projected position. 24 µm and FUV flux densities (hereafter $F_{24\mu m}$ and $F_{FUV}$, respectively) are shown in Figures 4-6.

## 3. Results and discussion

### 3.1 *Moment maps*
As shown in Figure 1, $^{12}$CO(1-0) gas exhibits a ring-like structure, i.e. there is a deficiency in CO gas in the central region, and the velocity field seems to be regular. However, as seen in the right panel of Figure 1, the velocity dispersion at the outskirts of the galaxy is higher (i.e. on the south-eastern and north-western sides of the disc). An increase in star formation activities, such as supernovae events, strong stellar winds from massive stars, turbulence and/or any interactions with nearby galaxies could cause an increase in the observed velocity dispersion [18 and references therein]. As the galaxy has not been reported to show any current interactions and /or collisions with other galaxies (NED, *HyperLEDA*), it is reasonable to assume that the level of star formation activity in the centre and outskirts of the galaxy could be different.

### 3.2. *Integrated intensity, gas mass and gas surface density*
Integrated CO line intensity, $M_{H_2}$ and $\sum H_2$ as a function of the projected distance from the galaxy's centre is shown in Figure 3. As seen from the figure all three parameters, i.e. CO line intensity, $M_{H_2}$ and $\sum H_2$, increase up to about 40 arcsec from the galaxy's centre and then start to decrease. The same behaviour is seen on each side of the disc (see the bottom panel of Figure 3). However, the most values on the eastern side of the disc are higher than those on the western side, i.e. higher median values on the eastern side (see blue and red dashed lines in the bottom panel of Figure 3). This indicates that the eastern side of the disc is slightly brighter in CO with higher $M_{H_2}$ and $\sum H_2$ values compared to the western side.

### 3.3. *$H_2$ mass, extinction, 24 µm and FUV flux densities*
$M_{H_2}$ as a function of the extinction (i.e. $F_{24\mu m}/F_{FUV}$ ratio), $F_{24\mu m}$ and $F_{FUV}$ are shown in Figure 4. There is a strong positive correlation between $M_{H_2}$ and the extinction (*Spearman correlation coefficient* of $r_s = 0.70$). The correlation is stronger on the western side of the disc (i.e. $r_{s(W)} = 0.86$) than the eastern side (i.e. $r_{s(E)} = 0.75$). As seen from panel *a* in Figure 4, a similar but weaker correlation exists between $M_{H_2}$ and $F_{24\mu m}$ (i.e. $r_s = 0.55$). However, the correlation between $M_{H_2}$ and $F_{FUV}$ is negative and weak ($r_s = -0.39$), i.e. $M_{H_2}$ decreases as $F_{FUV}$ increases (see panel *b* in Figure 4). Overall, $M_{H_2}$ shows a positive correlation with the extinction and $F_{24\mu m}$, while the correlation between $M_{H_2}$ and $F_{FUV}$ is negative and weaker.

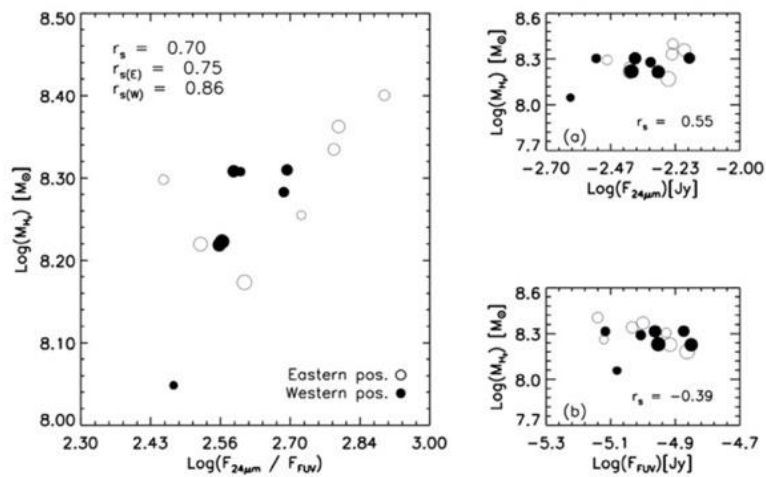

Figure 4. The molecular gas mass, $M_{H_2}$, as a function of the extinction (i.e. $F_{24\mu m}/F_{FUV}$ ratio), 24 μm flux density (panel a in the top right) and FUV flux density (panel b in the bottom right) are shown. Open circles represent the projected positions on the eastern side of the disc while the black filled circles indicate those on the western side. In all panels, the size of the symbols was arranged so that the smallest symbol represents the farthest projected position from the centre while the largest symbol indicates the closest projected position to the centre. *Spearman correlation coefficient* ($r_s$) is shown in each panel. In the left panel, $r_{s(E)}$ and $r_{s(W)}$ represent the values for the projected positions on the eastern and western side of the disc, respectively.

As seen from Figure 5, the extinction shows a similar variation to that of intensity, gas mass and gas surface density (see Section 3.2 and Figure 3), i.e. from the centre through the outskirts, the extinction increases up to about 40 arcsec and then starts to decrease (see panels *a* and *b* in Figure 5). The extinction also shows higher values on the eastern side of the disc (see open circles in panels *a* and *b* in Figure 5) similar to the intensity, gas mass and surface density (see Figure 3).

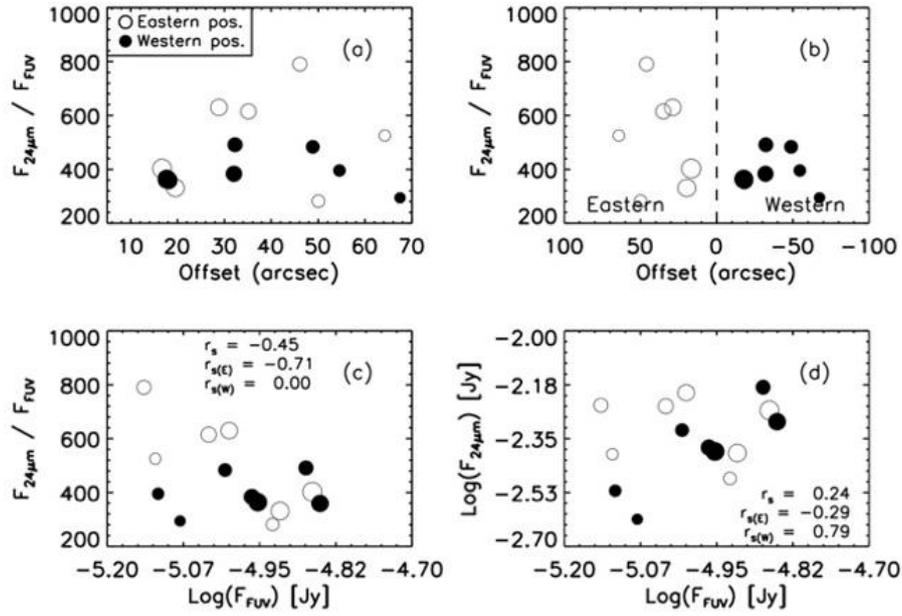

Figure 5. The extinction (i.e. ratio of $F_{24\mu m}/F_{FUV}$) as a function of the projected distance (panel a), the distance on each side of the disc (panel b) and FUV flux (panel c) are shown. $F_{24\mu m}$ as a function of $F_{FUV}$ is shown in panel d. The size of the symbols was arranged as stated in the caption of Figure 4. *Spearman correlation coefficient* ($r_s$) is shown on the lower panels.

As seen from panel *c* of Figure 5, there is a strong negative correlation between the extinction and $F_{FUV}$ on the eastern side of the disc, i.e. $F_{FUV}$ increases as the extinction decreases (i.e. $r_{s(E)} = -0.71$). However, no statistically meaningful correlation was found between the two parameters on the western side (i.e. $r_{s(W)} = 0$). As indicated by the panel *d* in Figure 5, interestingly, although overall trend indicates a weak positive correlation between $F_{24\mu m}$ and $F_{FUV}$ ($r_s = 0.24$), the situation is different on each side of the disc, i.e. there is a weak negative correlation between $F_{24\mu m}$ and $F_{FUV}$ on the eastern side (i.e. $r_{s(E)} = -0.29$) while the correlation is much stronger and positive on the western side (i.e. $r_{s(W)} = 0.79$).

As seen from the top panel in Figure 6, from the centre up to about 40 arcsec on each side of the disc, $F_{24\mu m}$ seems to increase and then starts to decrease. However, $F_{FUV}$ gradually decreases as a function of the projected distance, and the correlation between $F_{FUV}$ and the distance is stronger on each side of the disc compared to the correlation between $F_{24\mu m}$ and the distance (see the lower panel in Figure 6).

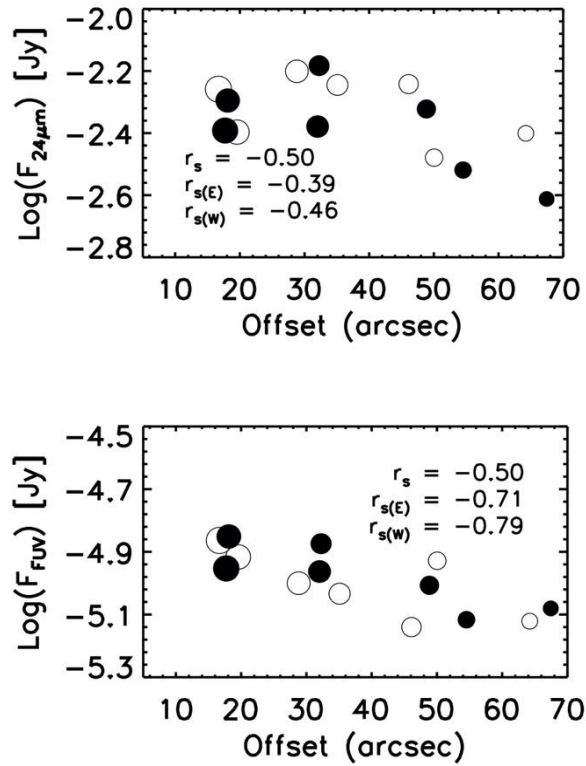

Figure 6. $F_{24\mu m}$ and $F_{FUV}$ as a function of the projected distance from the galaxy's centre are shown. Black filled circles indicate the projected positions on the western side, while open circles indicate the positions on the eastern side.

## 4. Conclusions

We probe the physical properties of the ISM at 14 projected positions covering the entire gaseous disc of the spiral galaxy NGC 7331. Based on our multi-wavelength analysis we summarise our main conclusions as below:

Moment maps revealing a ring-like gas distribution indicate a central deficiency in CO, and CO is brighter in the eastern side of the gaseous disc. The gas kinematics of the galaxy is generally relaxed (i.e. velocity field is mostly regular), but the velocity dispersion at the outskirts seems to be higher.

Integrated intensity, molecular gas mass, gas surface density, the extinction and 24 μm dust emission increase up to about 40 arcsec and then start to decrease. However, FUV flux densities gradually decrease (i.e. without showing a prominent peak) as a function of the projected distance from the galactic centre. Values for all parameters on the eastern side of the disc tend to be higher than those on the western side (except FUV flux).

There is a strong positive correlation between $M_{H_2}$, and the extinction (i.e. $r_s = 0.70$) over the disc (stronger in the western side of the disc, i.e. $r_{s(W)} = 0.86$ ). However, there is a relatively weaker positive correlation between $M_{H_2}$ and 24 μm flux density, $F_{24\mu m}$, (i.e. $r_s = 0.55$) and a weaker negative correlation between $M_{H_2}$ and FUV flux density, $F_{\text{FUV}}$, (i.e. $r_s = -0.39$).

There is a negative correlation between $F_{\text{FUV}}$ and the extinction on the eastern side of the disc, while no correlation found between the two on the western side. Although there is a positive correlation between $F_{24\mu m}$ and $F_{\text{FUV}}$ on the western side, the correlation is much weaker and negative on the eastern side. This indicates that $F_{\text{FUV}}$ is increasing at a higher rate on the eastern side as the $F_{24\mu m}$ is slowly decreasing. However, on the western side, the increase in both $F_{24\mu m}$ and $F_{\text{FUV}}$ seem to be happening at a similar rate, causing the flat correlation between the extinction and $F_{\text{FUV}}$ seen on the western side.

In summary, our results indicate that the ISM on each side of the disc has some differences in terms of $F_{\text{FUV}}$ and its correlations with other parameters, while integrated CO line intensity, $M_{H_2}$, $\sum H_2$, $F_{24\mu m}$ and the extinction indicate a similar distribution over the disc (i.e. there is a 'bump' around 40 arcsec from the centre on each side of the disc, and the values tend to be higher on the eastern side).

### Acknowledgements

This work is based [in part] on observations made with the Spitzer Space Telescope, which was operated by the Jet Propulsion Laboratory, California Institute of Technology under a contract with NASA. We acknowledge the usage of the *HyperLEDA* database (http://leda.univ-lyon1.fr). This research has made use of the NASA/IPAC Infrared Science Archive, which is operated by the Jet Propulsion Laboratory, California Institute of Technology, under contract with the National Aeronautics and Space Administration.

**Appendix**

**Appendix A.** Spectrum for each studied position.

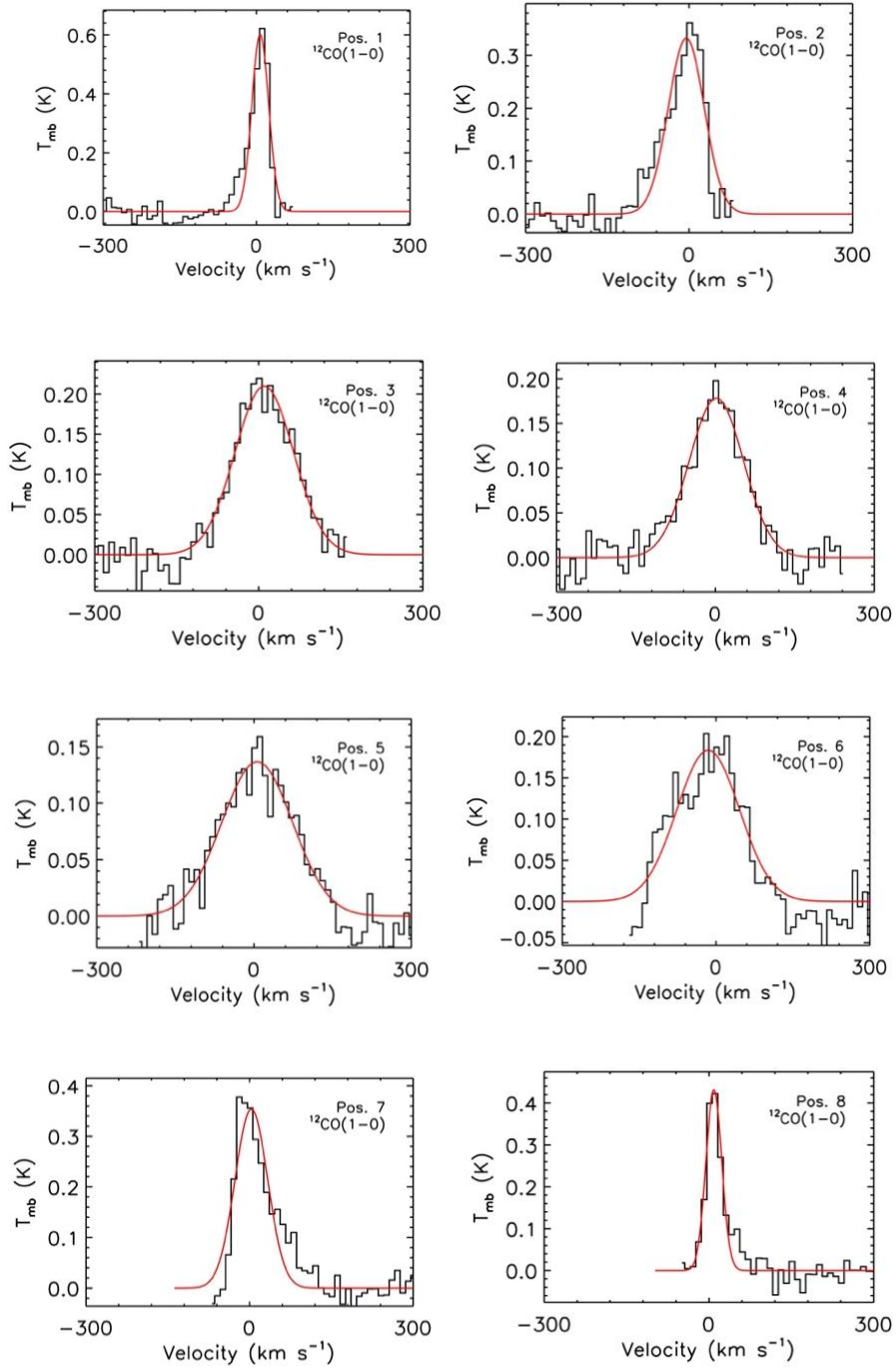

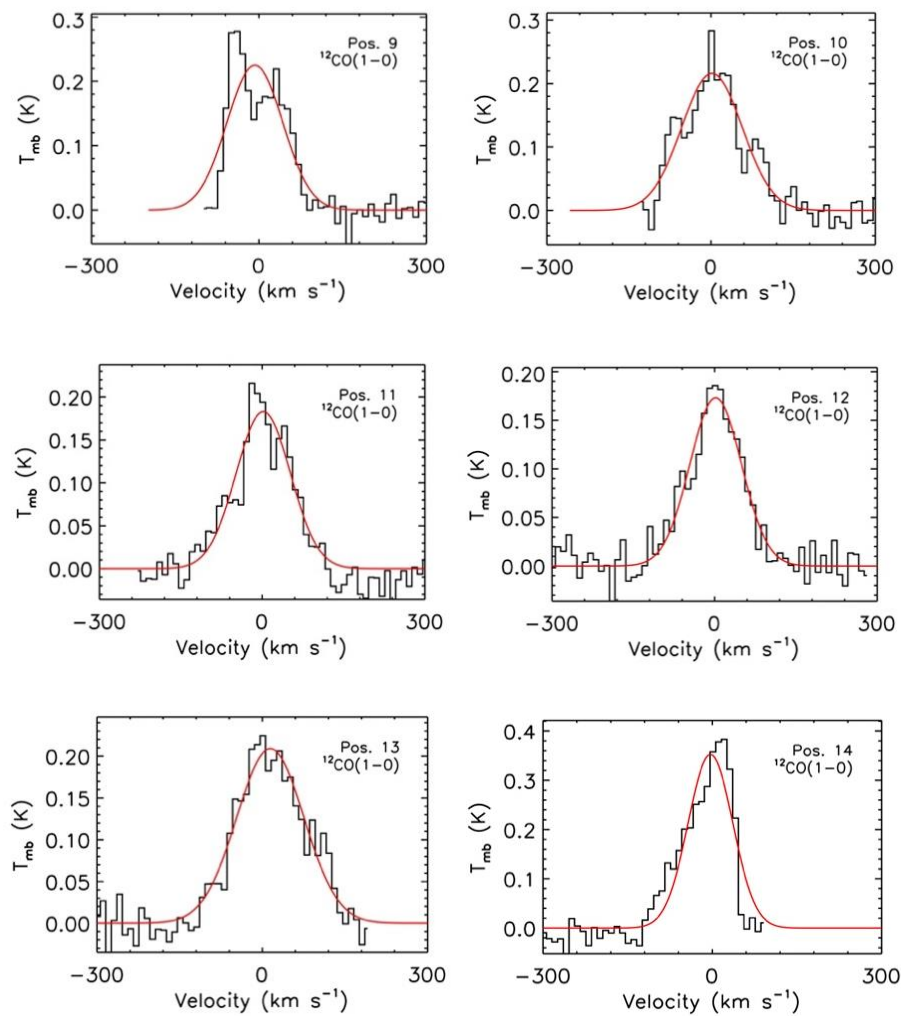

Figure A.1 The CO spectra extracted from the projected positions are shown. The red line indicates the best-fit single Gaussian profile. The position numbers (also shown in Figure 2) are indicated in the upper right corner of each spectrum.